\documentclass[usenatbib]{mn2e}

\usepackage{amsmath}
\usepackage{eufrak}
\usepackage{graphicx}
\usepackage[colorlinks=true, linkcolor=blue, citecolor=blue, urlcolor=blue]{hyperref}
\usepackage{times}
\pdfoutput=1

\newcommand{\bfr}{{\boldsymbol{r}}}
\newcommand{\bfx}{{\boldsymbol{x}}}
\newcommand{\noise}{{\mathcal{R}}}
\newcommand{\txd}{{\text{d}}}

\begin{document}

\title{Smart detectors for Monte Carlo radiative transfer}

\author[M.~Baes]{Maarten Baes \\ Sterrenkundig Observatorium,
Universiteit Gent, Krijgslaan 281-S9, B-9000 Gent, Belgium, maarten.baes@ugent.be}

\maketitle

\begin{abstract}
  Many optimization techniques have been invented to reduce
  the noise that is inherent in Monte Carlo radiative transfer
  simulations. As the typical detectors used in Monte Carlo
  simulations do not take into account all the information contained
  in the impacting photon packages, there is still room to optimize
  this detection process and the corresponding estimate of the surface
  brightness distributions. We want to investigate how all the
  information contained in the distribution of impacting photon
  packages can be optimally used to decrease the noise in the surface
  brightness distributions and hence to increase the efficiency of
  Monte Carlo radiative transfer simulations. 

  We demonstrate that the estimate of the surface brightness
  distribution in a Monte Carlo radiative transfer simulation is
  similar to the estimate of the density distribution in an SPH
  simulation. Based on this similarity, a recipe is constructed for
  smart detectors that take full advantage of the exact location of
  the impact of the photon packages. Several types of smart detectors,
  each corresponding to a different smoothing kernel, are
  presented. We show that smart detectors, while preserving the same
  effective resolution, reduce the noise in the surface brightness
  distributions compared to the classical detectors. The most
  efficient smart detector realizes a noise reduction of about 10\%,
  which corresponds to a reduction of the required number of photon
  packages (i.e.\ a reduction of the simulation run time) of 20\%. As
  the practical implementation of the smart detectors is
  straightforward and the additional computational cost is completely
  negligible, we recommend the use of smart detectors in Monte Carlo
  radiative transfer simulations.
\end{abstract}


\section{Introduction}
\label{introduction.sec}

The Monte Carlo method \citep[e.g.][]{1959CashEver,
  1996ApJ...465..127B, 2001ApJ...551..269G, 2003MNRAS.343.1081B,
  2003A&A...399..703N} has become one of the most popular methods to
perform radiative transfer simulations. One of the greatest advantages
of the Monte Carlo method is its conceptual simplicity. Instead of
solving the radiative transfer equations that describe the radiation
field, Monte Carlo simulations actually follow the photon packages
that make up the radiation field in a very natural way. This ensures
that, compared to other approaches to multi-dimensional radiative
transfer problems, the practical implementation of Monte Carlo
radiative transfer is surprisingly straightforward. This simple,
straightforward approach also enables the inclusion of additional
ingredients, such as the polarization of scattered light
\citep{1995ApJ...441..400C, 1996ApJ...465..127B}, the kinematics of
the sinks and sources \citep{2001ApJ...548..150M, 2002MNRAS.335..441B}
and gas ionization and recombination processes
\citep{2003MNRAS.340.1136E, 2005MNRAS.362.1038E}. On the other hand,
probably the most important disadvantage of the Monte Carlo method is
the appearance of Poisson noise, which is inherently tied to the
probabilistic nature of the method. In pure Monte Carlo radiative
transfer simulations, the noise in any observed property goes as
$1/\sqrt{N}$ with $N$ the number of photon packages, which means that
one needs to quadruple the number of photon packages to halve the
errors. Due to this slow convergence, radiative transfer simulations
based on the most simple application of the Monte Carlo method are
very inefficient and have difficulties to compete with other methods.

Fortunately, in the many years since the first applications of the
Monte Carlo method to radiative transfer simulations, several
intelligent optimization techniques have been invented to increase the
efficiency of the Monte Carlo techniques. These techniques, equivalent
to so-called variance reduction techniques in Monte Carlo integration,
aim at reducing the noise by including deterministic elements into the
probabilistic simulation. One of the first examples of such techniques
is the well-known forced first scattering technique, already included
in the first implementations of Monte Carlo radiative transfer
\citep{1959CashEver, 1970A&A.....9...53M}. Other noise-reduction
techniques that have strongly increased the efficiency of the Monte
Carlo method include the use of weighted photon packages
\citep{1977ApJS...35....1W}, the peel-off technique
\citep{1984ApJ...278..186Y}, the treatment of absorption as a
continuous rather than a discrete process \citep{1999A&A...344..282L},
the frequency distribution adjustment technique
\citep{2001ApJ...554..615B, 2005NewA...10..523B} and the use of
polychromatic photon packages \citep{2005AIPC..761...27B,
  2006MNRAS.372....2J}.

One aspect of Monte Carlo radiative transfer where no significant
noise reduction techniques have been presented is the last step in the
life cycle of the photon packages, namely their detection. The goal of
most Monte Carlo radiative transfer simulations is to determine the
observed surface brightness distribution at some observer's
position. To construct the observed surface brightness distribution,
some kind of detector must be simulated, on which the photon packages
that leave the system are recorded. In the spirit that Monte Carlo
radiative transfer simulations mimic the real physical processes as
naturally as possible, the simulated detectors are usually natural
(idealized) representations of actual CCD detectors. They basically
consist of a two-dimensional array of pixels, which act as a reservoir
for the incoming photon packages. When a photon package leaves the
system and arrives at the location of the observer, the correct bin is
determined and the luminosity recorded in the bin is increased with
the luminosity of the photon package. At the end of the simulation,
the detector is read out like a CCD detector and the surface
brightness distribution is constructed.

While this approach seems the most natural way to simulate the
detection of photon packages in a Monte Carlo simulation, it might not
be the most efficient. We must be aware that, although we are
simulating a real detection as closely as possible, we have
{\em{more}} information at our disposal than real observers. The
maximum information that a real observer can obtain (in the academic
limit of perfect noise-free observations and instruments) when imaging
with a CCD detector, is the number of photon packages that arrive in
each of his pixels. As numerical simulators, we have at our disposal
the full information on the precise location of the impact of each
photon package on the detector. It would be a pity to throw away this
useful additional information just in order to mimic the behaviour of
a real CCD detector.

The goal of this paper is to investigate how this additional
information can be used to decrease the noise in the estimated surface
brightness distributions and hence to increase the efficiency of Monte
Carlo radiative transfer simulations. In Section~2 we describe the
classical way of detecting photon packages as a smoothing process,
similar to the averaging processes encountered in the smoothed
particle hydrodynamics framework for hydrodynamical simulations. We
use this similarity between both processes to construct a new set of
{\em{smart}} detectors, which aim at curing two drawbacks of the
classical detectors. In Section~3 we test the accuracy and
performance of these smart detectors and compare them with the
classical detector. The results are summarized and discussed in
Section~4.

\section{Smart detectors}

\subsection{The classical detector}

A typical detector in Monte Carlo simulations is based on a realistic
CCD detector. It consists of a rectangular two-dimensional array of
bins of dimension $\Delta$ placed on the plane of the sky. We denote
the centre positions of these bins as $\bar\bfx_{ij} =
(\bar{x}_i,\bar{y}_j)$,
\begin{gather}
  \bar{x}_i = \bar{x}_{\text{min}} + i\,\Delta,
  \\
  \bar{y}_j = \bar{y}_{\text{min}} + j\,\Delta.
\end{gather}
When the $k$'th photon package hits the detector at a certain position
$\bfx_k=(x_k,y_k)$, we find out in which bin it will end,
and we add the luminosity $L_k$ carried by photon package to the
number of previously detected photon packages in this particular bin. At the
end of the simulation, we determine our estimate
$I_{\text{s}}(\bar\bfx_{ij})$ for the surface brightness at the
position $\bar\bfx_{ij}$ by summing the contribution of all photon packages
that have been recorded in that bin and correcting for the surface of
the bin,
\begin{equation}
  I_{\text{s}}(\bar\bfx_{ij})
  =
  \frac{1}{\Delta^2}
  \sum_{k=1}^N
  L_k\,N_{ij}(\bfx_k)
  \label{Isfirstformula}
\end{equation}
where 
\begin{equation}
  N_{ij}(\bfx)
  =
  \begin{cases}
    \;1
    &\quad
    \text{if $\bar{x}_i-\tfrac12\Delta\leq x\leq \bar{x}_i+\frac12\Delta$}
    \\
    &\qquad
    \text{and $\bar{y}_j-\tfrac12\Delta\leq y\leq \bar{y}_j+\frac12\Delta$},
    \\[1em]
    \;0
    &\quad
    \text{else}.
  \end{cases}
\end{equation}

\subsection{The link to SPH}

For an interesting point of view on simulating the detection of
photon packages and the calculation of the surface brightness distribution, we
now shift to another important technique in computational
astrophysics, namely smoothed particle hydrodynamics or SPH
\citep{1977MNRAS.181..375G, 1977AJ.....82.1013L}. SPH is a
computational technique for hydrodynamical simulations in which the a
fluid is represented as a finite collection of fluid elements. To
estimate the value of a physical field $f(\bfr)$ at an arbitrary
position $\bfr$, we must perform local averages over volumes of
nonzero extent (i.e.\ over a finite number of fluid particles). The
mean or smoothed value of $f(\bfr)$, denoted as $f_{\text{s}}(\bfr)$
can be determined through kernel estimation,
\begin{equation}
  f_{\text{s}}(\bfr)
  =
  \iiint W(\bfr-\bfr')\,f(\bfr')\,\txd\bfr',
\label{SPHbasic}
\end{equation}
where $W(\bfr)$ is the so-called smoothing kernel, which should be
normalized according to
\begin{equation}
  \iiint W(\bfr)\,\txd\bfr = 1,
\end{equation}
and which typically is strongly peaked about $\bfr=0$. In particular,
if the fluid mass distribution $\rho(\bfr)$ is represented by a flow
of $N$ particles with mass $m_k$, we have
\begin{equation}
  \rho(\bfr)
  =
  \sum_{k=1}^N m_k\,\delta(\bfr-\bfr_k),
\end{equation}
and the estimate $\rho_{\text{s}}(\bfr)$ for the mass distribution is
given by
\begin{equation}
  \rho_{\text{s}}(\bfr)
  =
  \sum_{k=1}^N m_k\,W(\bfr-\bfr_k).
\label{SPHmd}
\end{equation}
The standard interpretation of this formula is to see the fluid
elements that make up the representation of the fluid not as
point-like particles, but as particles with a smoothed smeared out
mass distribution $\rho_k(\bfr) = m_k\,W(\bfr-\bfr_k)$. The total
density at any position $\bfr$ is then the sum of the contribution of
the $N$ particles in the fluid.

Returning back to our simulated CCD detector, we note that we can
write expression~(\ref{Isfirstformula}) as
\begin{equation}
  I_{\text{s}}(\bar\bfx_{ij})
  =
  \sum_{k=1}^N
  L_k\,W(\bar\bfx_{ij}-\bfx_k),
\label{Is-discr}
\end{equation}
with $W(\bfx)$ a function defined as
\begin{multline}
  W(\bfx)
  =
  \left[
    \frac{H\left(x+\tfrac12\Delta\right)
      -
      H\left(x-\tfrac12\Delta\right)}{\Delta}
  \right]\,
  \\
  \times
  \left[
    \frac{H\left(y+\tfrac12\Delta\right)
      -
      H\left(y-\tfrac12\Delta\right)}{\Delta}
  \right]\,
\label{Wheavi}
\end{multline}
with $H$ the Heaviside step function. Comparing
equations~(\ref{Is-discr}) and (\ref{SPHmd}) we immediately see a
connection between the computation of the (three-dimensional) mass
density in fluids in the SPH formalism and the computation of the
(two-dimensional) surface brightness in a Monte Carlo radiative
transfer simulation. We can make this connection clearer by rewriting
equation~(\ref{Is-discr}) as
\begin{equation}
  I_{\text{s}}(\bar\bfx_{ij})
  =
  \iint W(\bar\bfx_{ij}-\bfx')\,I(\bfx')\,\txd\bfx',
\label{Is-cont}
\end{equation}
with
\begin{equation}
  I(\bfx)
  =
  \sum_{k=1}^N L_k\,\delta(\bfx-\bfx_k).
\end{equation}
Since this latter expression is nothing but the ``true'' surface
brightness distribution corresponding to $N$ photon packages hitting the
detector plane (each photon package results in a Dirac delta function),
equation~(\ref{Is-cont}) is the direct analogue of the SPH basic
equation~(\ref{SPHbasic}). The bottom-line is that we can interpret the
radiation field at the plane of the sky as a fluid, which is
represented by a finite number of smoothed photon packages. Each smoothed
photon package corresponds to a smeared out surface brightness distribution
$I_k(\bfx) = L_k\,W(\bfx-\bfx_k)$. Formula~(\ref{Is-discr}) shows that
the total observed surface brightness distribution at the positions
$\bfx_{ij}$ is the sum of the contributions of each of these photon packages.

\subsection{Smart detectors}

We have seen that we can interpret the determination of the surface
brightness distribution in a Monte Carlo radiative transfer simulation
as a smoothing operation similar to the averaging in SPH
hydrodynamical simulations. As in SPH simulations, we can consider
using a different kernel --- in the present case this corresponds to a
different kind of detector. In principle, there is no restriction on
the shape of the function $W(\bfx)$ apart from the normalization
condition
\begin{equation}
  \iint W(\bfx)\,\txd\bfx = 1,
\end{equation}
and the requirement that $W(\bfx)$ is a centrally peaked
function. This freedom allows us to construct {\em{smart}} detectors
which improve upon a number of potential disadvantages of the
traditional detector.

A first drawback of the smoothing kernel~(\ref{Wheavi}) is that it is
not isotropic, meaning that it has a preferential direction. Not all
photon packages hitting the detector at the same given distance from a grid
point $\bar\bfx_{ij}$ have the same impact on the surface
brightness. For example, a photon package hitting the detector at
\begin{equation}
  \bfx_k 
  =
  \left(\bar{x}_i+\tfrac23\,\Delta,\bar{y}_j\right)
\end{equation} 
will not contribute at all to the surface brightness
$I_{\text{s}}(\bfx_{ij})$, whereas a photon package hitting the detector at
\begin{equation}
  \bfx_k 
  = 
  \left(\bar{x}_i+\sqrt{\tfrac23}\,\Delta,
  \bar{y}_j+\sqrt{\tfrac23}\,\Delta\right),
\end{equation} 
at the same distance, will fully contribute to the estimate of the
surface brightness at this grid point. It is obvious that this problem
is solved when we consider circularly symmetric smoothing kernels
$W(\bfx)\equiv W(R)$. 

A second drawback of the classical detector is that it does not take
into account all the information that is contained in the impacting
photon package. For every photon package that falls onto our detector, we know the
exact location of the impact. The only information that a classical
simulated detector uses is the bin in which this location falls,
without any discrimination of the exact location within this bin. As
argued in the introduction, it would be a pity to not use this
information just for the sake of simulating a real CCD detector. To
solve the second problem, we need to look for kernels that give more
weight to impacts very close to the grid points than to impacts at
larger distance, which means that we need a kernel $W(R)$ that is a
monotonically decreasing function of $R$.

These are just minor limitations and still leave a lot of room for
different smart detectors. We can inspire ourselves on the smoothing
kernels that are often used in SPH simulations. The prototypical
normalized kernel is a gaussian, in two dimensions,
\begin{equation}
  W(R)
  =
  \frac{1}{\pi h^2}\,\exp\left(-\frac{R^2}{h^2}\right).
\label{gaussiankernel}
\end{equation}
The parameter $h$ in this expression, and in all other kernels we will
discuss, is called the smoothing length. It gives the width of the
area over which the smoothing is effective (we will later determine
the optimal value for this parameter). An important drawback of the
gaussian kernel is its infinite support, meaning that every photon package
impacting on the detector has a finite contribution to the surface
brightness at all grid points. Every photon package hence in principle
requires a summation over all grid points, at most of which it
contributes an absolutely negligible contribution. 

These efficiency problems are resolved by introducing smoothing
kernels with a compact support. Several compact support kernels have
been used in the SPH literature, the most popular of them being
kernels based on B-splines. The $M_3$ spline kernel
\citep{1981csup.book.....H} in two dimensions takes the form
\begin{equation}
  W(R)
  =
  \frac{16}{13\pi h^2}
  \times
  \begin{cases}
    \;\frac32-2\,u^2
    &\quad
    \text{if $0\leq u\leq \tfrac12$},
    \\[0.5em]
    \;(\tfrac32-u)^2
    &\quad
    \text{if $\tfrac12\leq u\leq\tfrac32$},
    \\[0.5em]
    \;0
    &\quad
    \text{else},
  \end{cases}
\label{M3splinekernel}
\end{equation}
with $u=R/h$. It has a continuous first derivative. A similar spline
kernel but with a continuous second derivative is the $M_4$ spline
smoothing kernel, introduced by \citet{1985A&A...149..135M}. Its
two-dimensional form is
\begin{equation}
  W(R)
  =
  \frac{10}{7\pi h^2}
  \times
  \begin{cases}
    \;1-\frac32\,u^2+\tfrac34\,u^3
    &\quad
    \text{if $0\leq u\leq 1$},
    \\[0.5em]
    \;\tfrac14\,(2-u)^3
    &\quad
    \text{if $1\leq u\leq 2$},
    \\[0.5em]
    \;0
    &\quad
    \text{else}.
  \end{cases}
\label{M4splinekernel}
\end{equation}
In principle, we have not requested that the kernel is everywhere
non-negative. Some SPH simulations have adopted so-called superkernels,
which are accurate to third or fourth order, but which necessarily
become negative in some part of the domain. The most famous example of
such kernels is the supergaussian kernel \citep{1982JCP....46..429G}
but examples with finite support have also been considered
\citep{1985JCP....60..253M, 1992ARA&A..30..543M,
  2000ANM....34..363C}. Using such superkernels could potentially
increase the accuracy of the smoothing process, but it can have
serious consequences when there is a sharp change in the surface
brightness: an undershoot occurs and the recorded surface brightness
may become negative. In order to avoid such situations, we stick to
positive kernels.

\subsection{Determination of the smoothing length}

Based on the three smoothing kernels~(\ref{gaussiankernel}),
(\ref{M3splinekernel}) and (\ref{M4splinekernel}) we can construct
three different kinds of smart detectors. Remains to determine which
value to take for the smoothing length $h$ for each of these smart
detectors. Obviously, $h$ should be large enough to make the smoothing
operation meaningful, whereas too large values of $h$ will tend to
over-smooth and wash away the details of the surface brightness
distribution. We want the smoothing lengths comparable with the pixel
scale, such that the noise in the images remains uncorrelated on a
pixel-by-pixel scale (as for a classical detector). Our determination
of the optimal smoothing length is derived from the requirement that
the effective area or ``resolution'' of the smoothing kernel should be
identical to the resolution of the classical detector.

\begin{figure}
\centering
\includegraphics[width=0.45\textwidth]{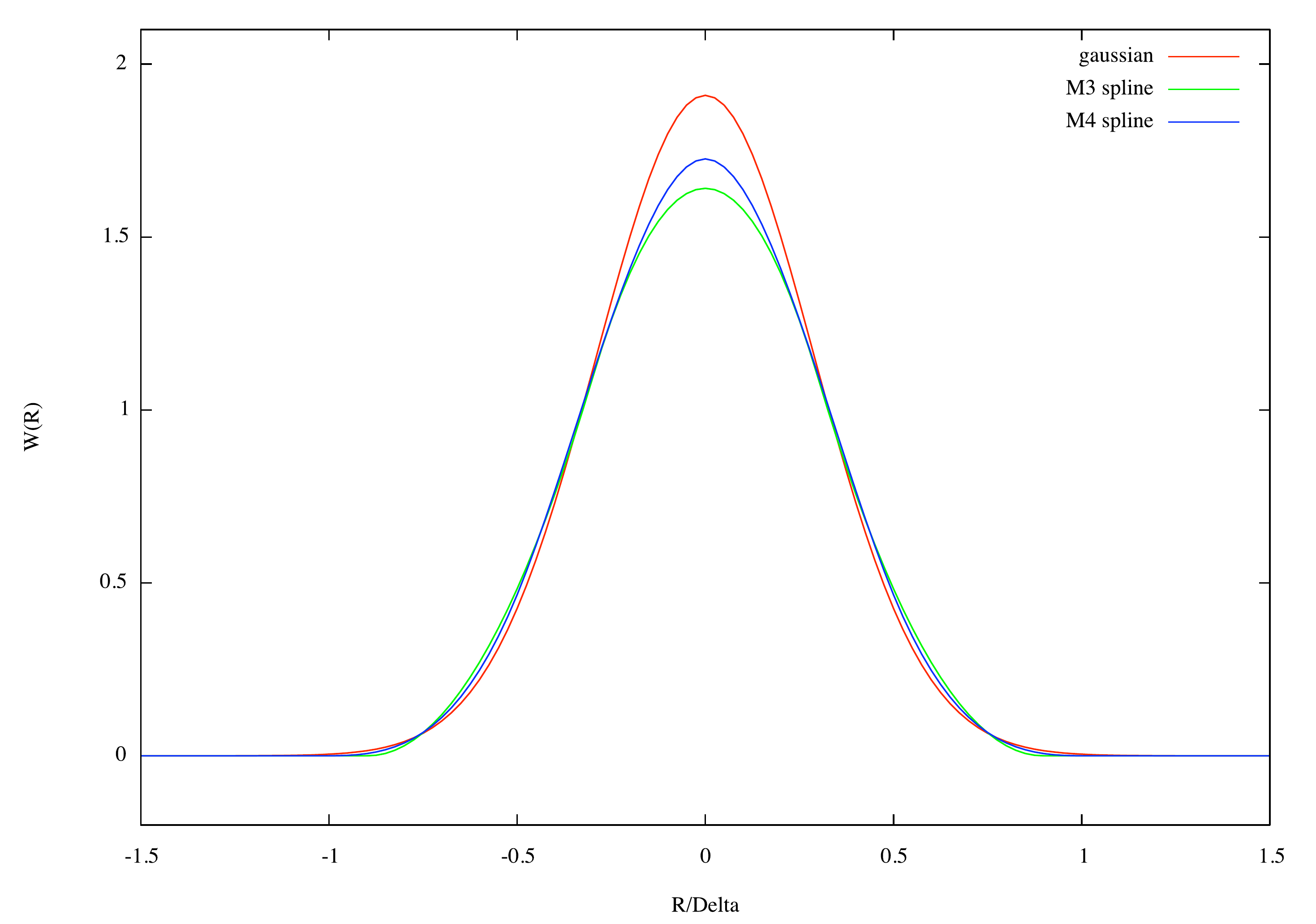}\hspace{2em}
\caption{The three smoothing kernels used for the construction of the
  smart detectors. The normalization and the width is the same for all
  three kernels.}
\label{kernels.pdf}
\end{figure}

There are various possibilities to identify the effective area of a
two-dimensional centrally peaked function. Probably the most general
one is the total dispersion $\sigma$, defined through
\begin{equation}
  \sigma^2
  =
  \iint W(\bfx)\,|\bfx|^2\,\txd\bfx.
\end{equation}
One can readily verify that the classical detector
kernel~(\ref{Wheavi}) has a dispersion
$\sigma=\Delta/\sqrt{6}$. Requiring that the resolution of the other
smoothing kernels are equal to this value we find as reference
smoothing lengths
\begin{align}
  h_{\text{ref}}
  &=
  \dfrac{1}{\sqrt{6}}\,\Delta
  \approx
  0.408\,\Delta
  &\text{(gaussian)},
  \\
  h_{\text{ref}}
  &=
  \dfrac{\sqrt{390}}{33}\,\Delta
  \approx
  0.598\,\Delta
  &\text{($M_3$ spline)},
  \\
  h_{\text{ref}}
  &=
  \dfrac{7}{\sqrt{186}}\,\Delta
  \approx
  0.513\,\Delta
  &\text{($M_4$ spline)}.
\label{href}
\end{align}
Figure~{\ref{kernels.pdf}} shows a comparison of the three different
smoothing kernels with their optimal smoothing lengths.

\section{Tests}

\subsection{Comparison of classical and smart detectors}

\begin{figure*}
\centering
\includegraphics[width=0.195\textwidth]{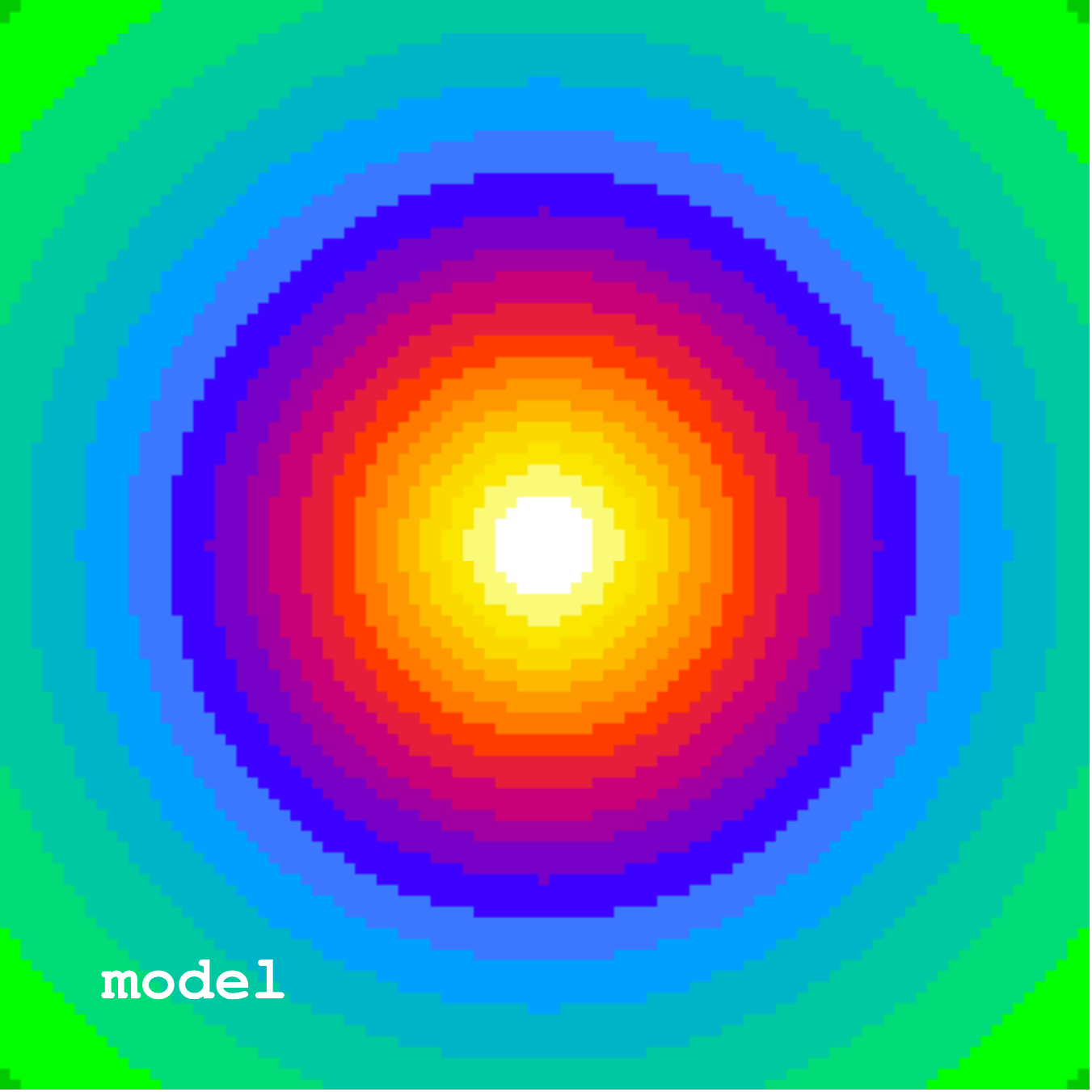}
\includegraphics[width=0.195\textwidth]{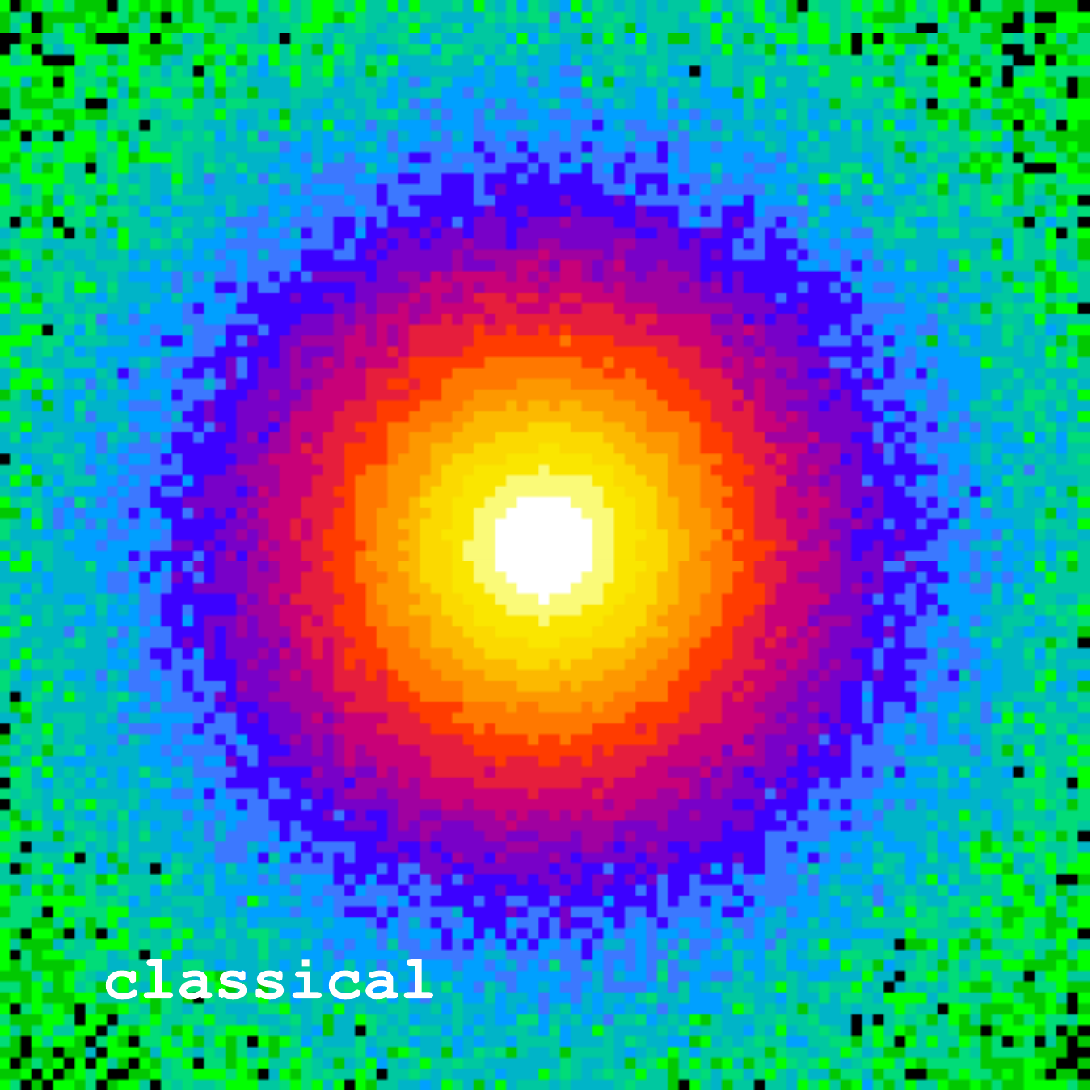}
\includegraphics[width=0.195\textwidth]{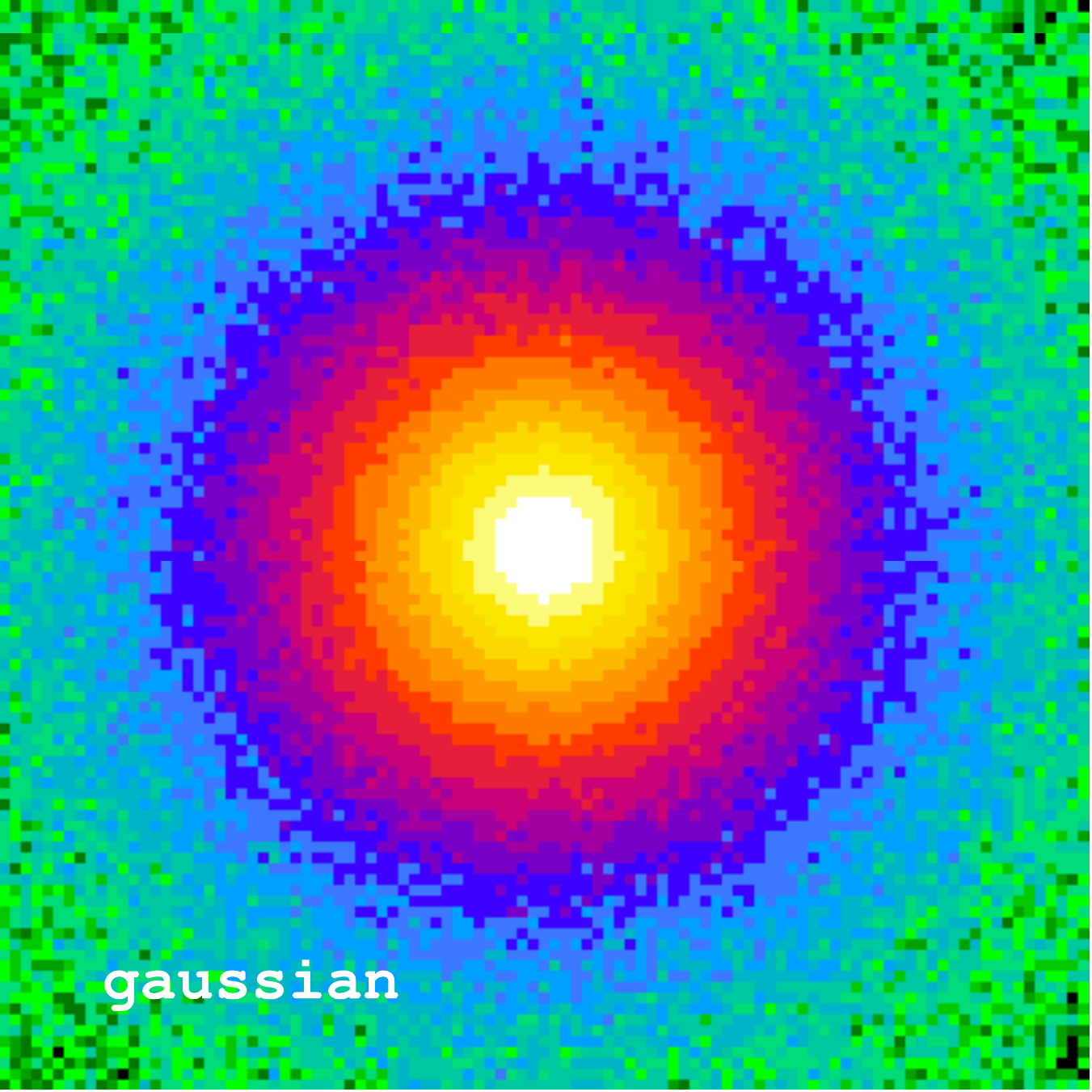}
\includegraphics[width=0.195\textwidth]{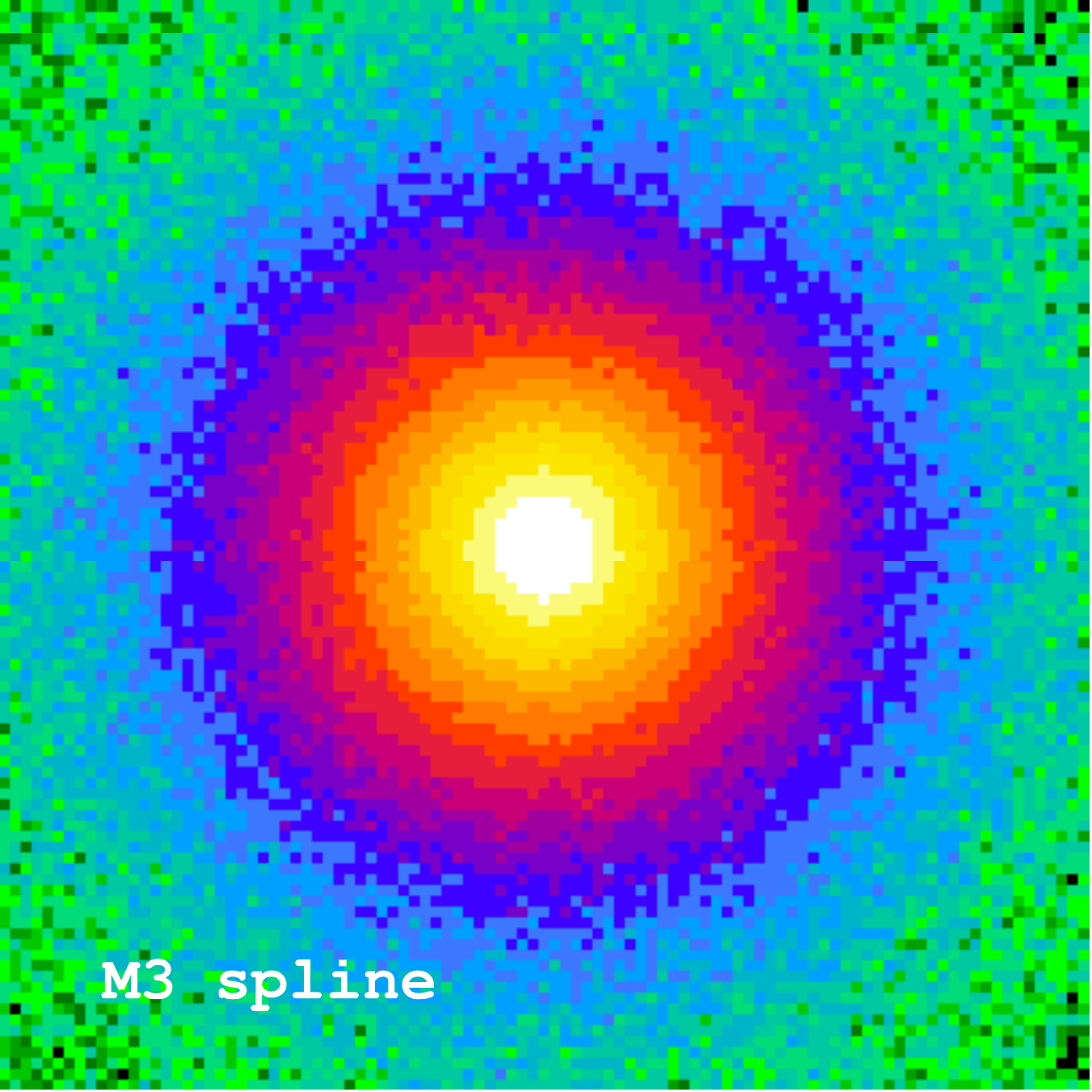}
\includegraphics[width=0.195\textwidth]{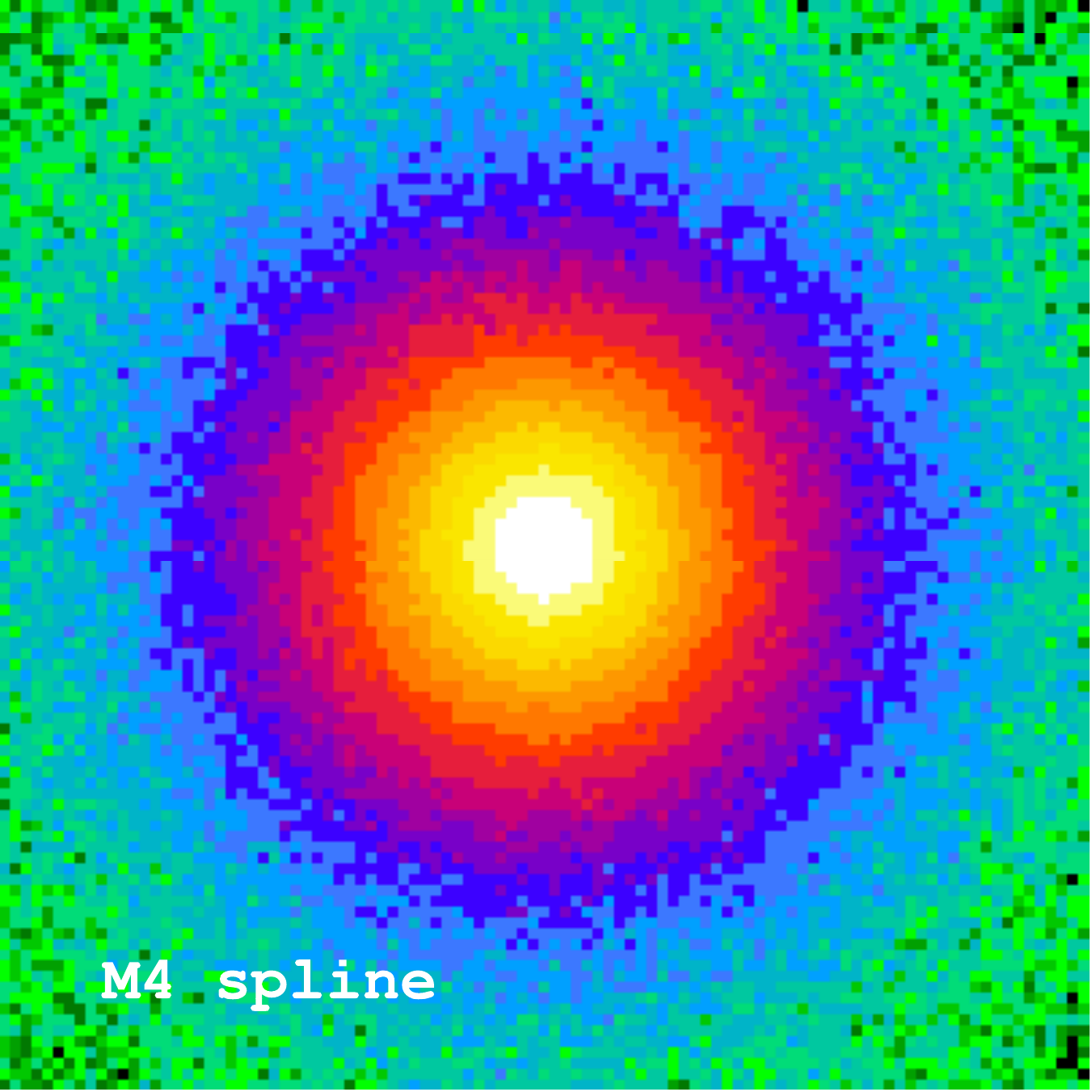} \\
\includegraphics[width=0.195\textwidth]{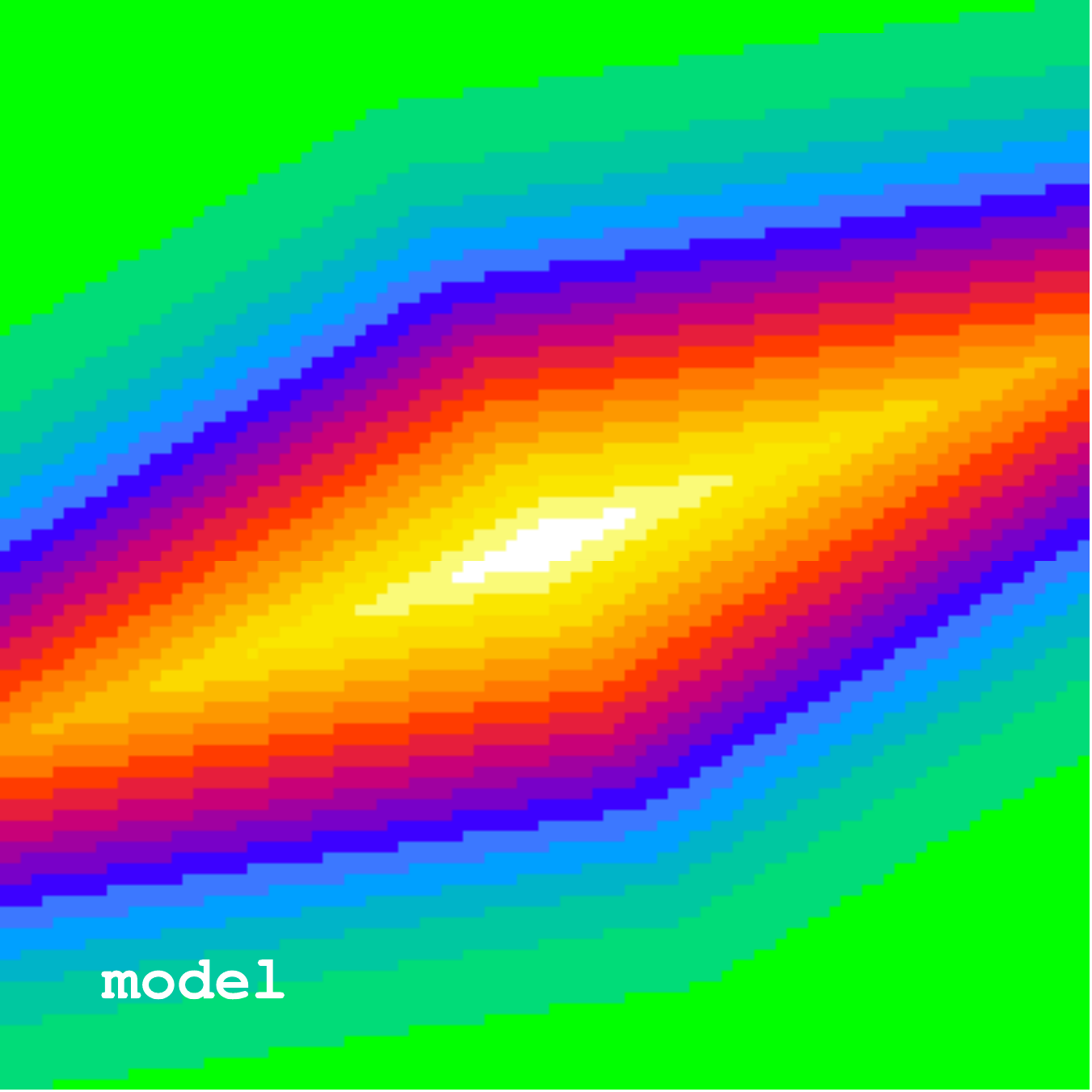}
\includegraphics[width=0.195\textwidth]{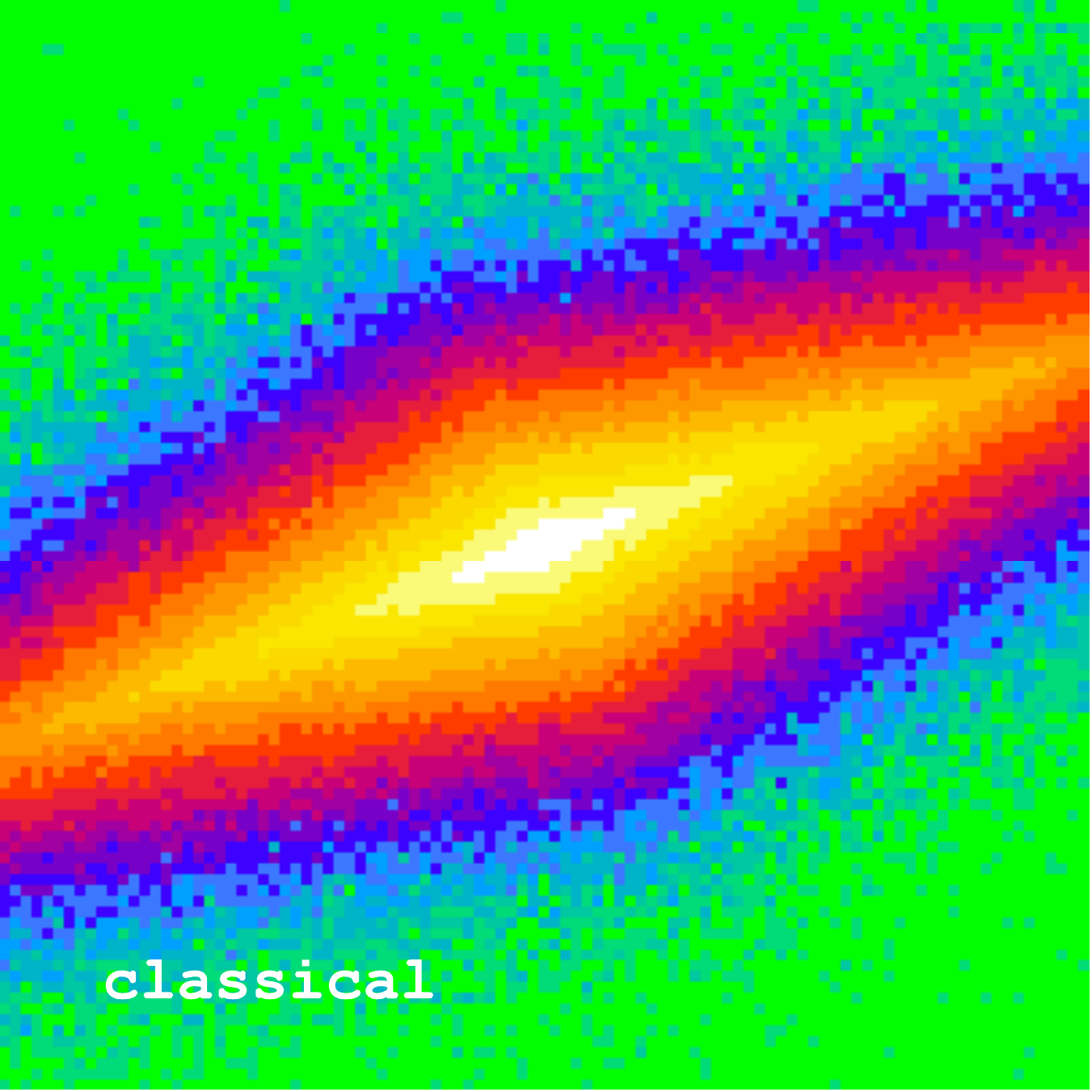}
\includegraphics[width=0.195\textwidth]{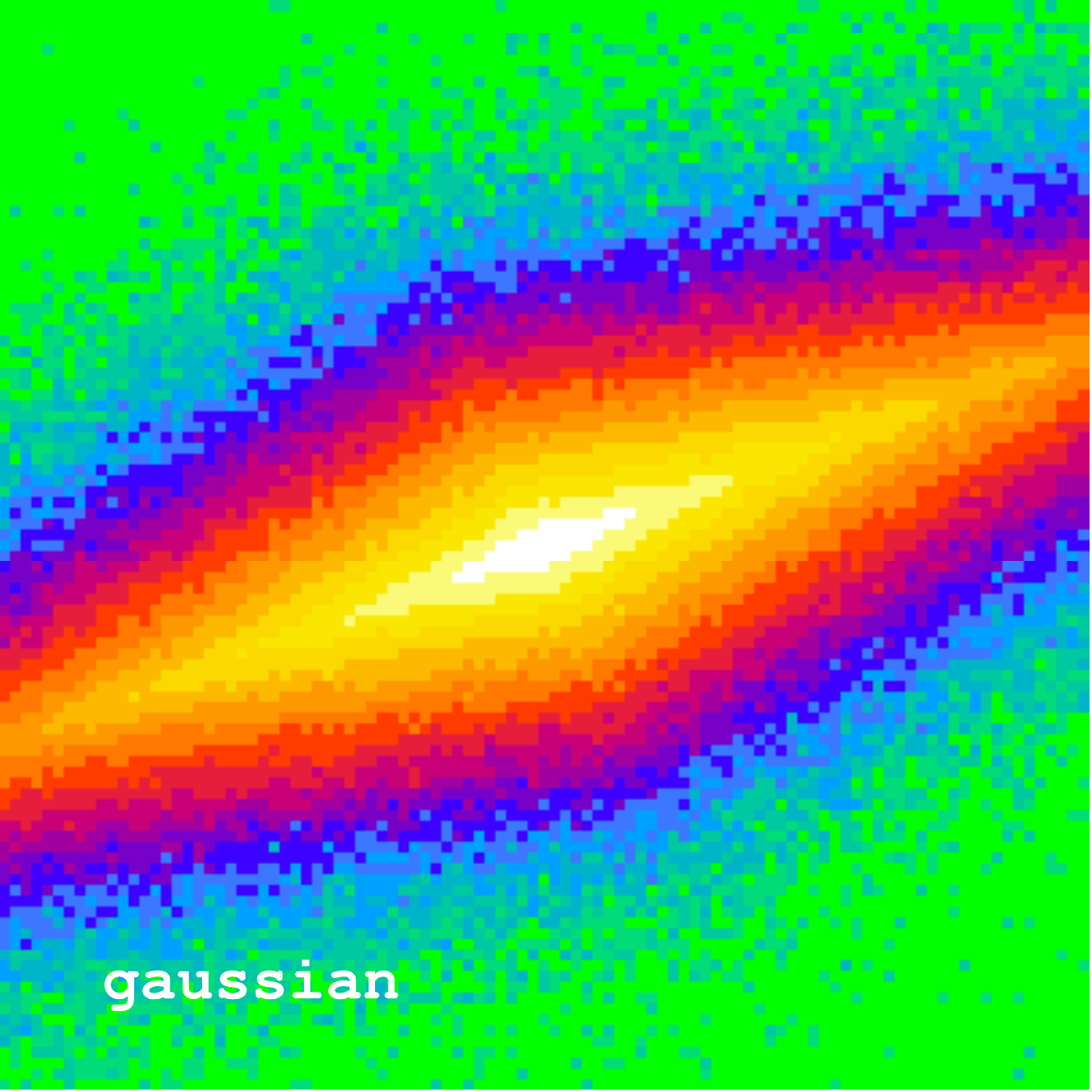}
\includegraphics[width=0.195\textwidth]{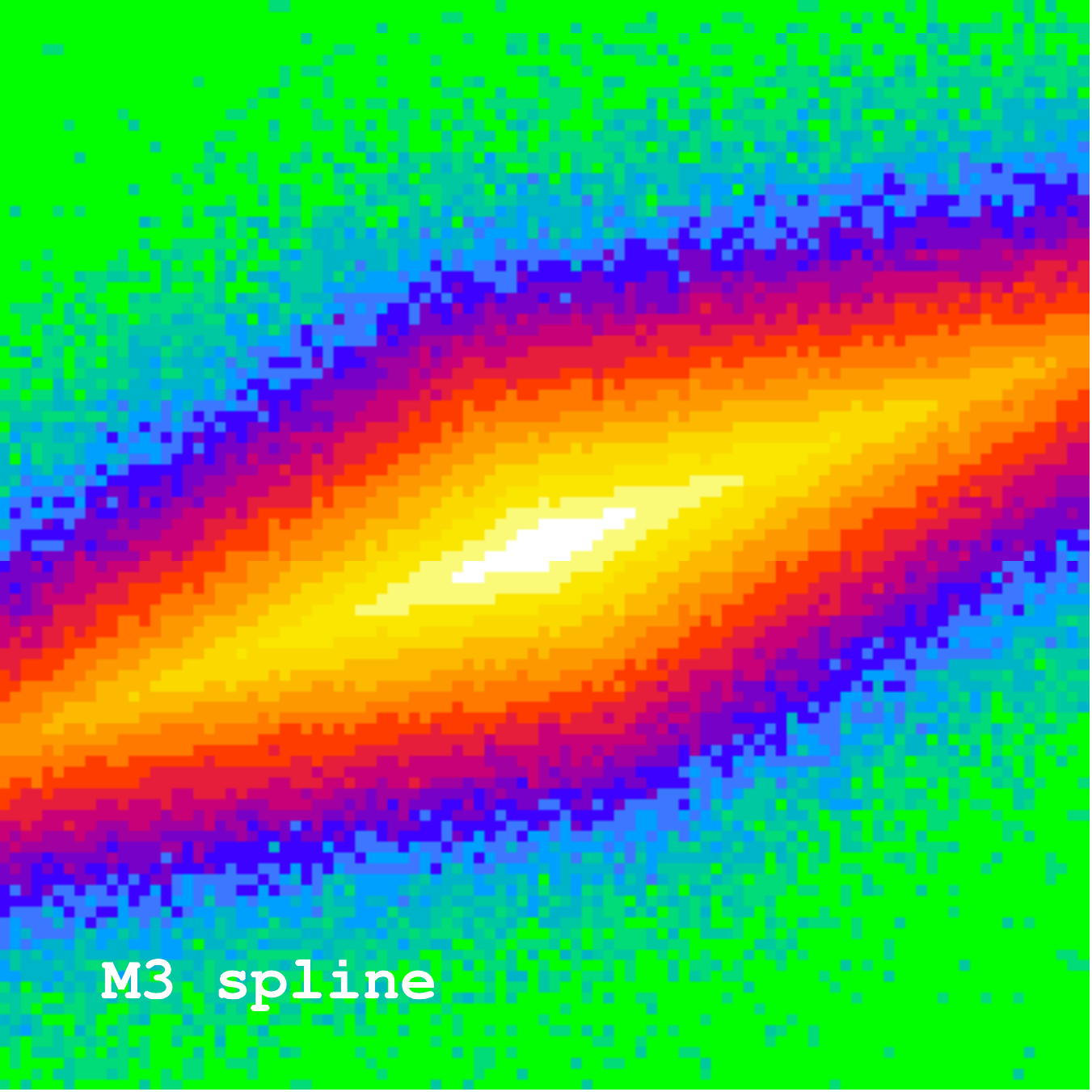}
\includegraphics[width=0.195\textwidth]{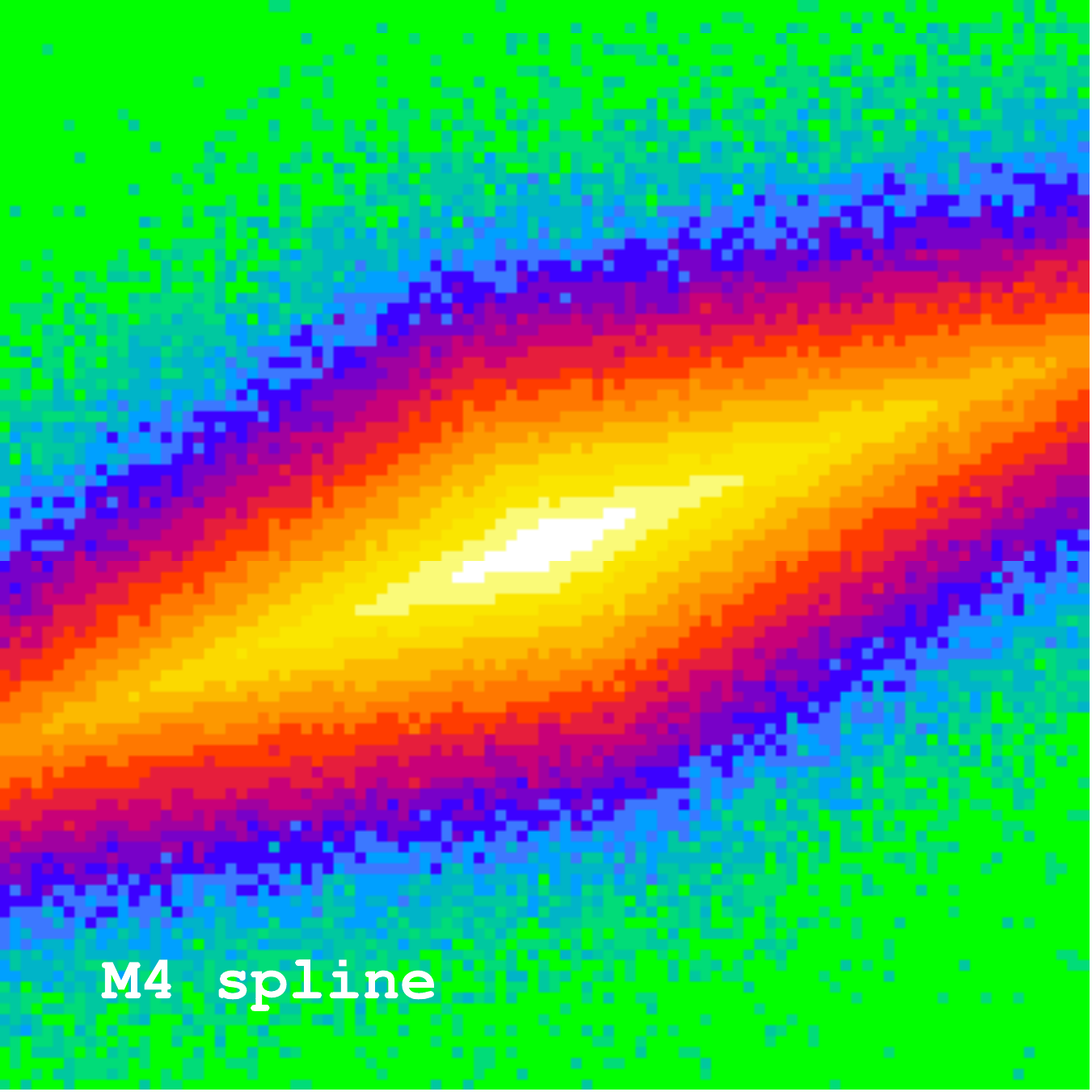}
\caption{The Plummer (top) and exponential disc (bottom) model surface
  brightness distribution $I(\bfx_{ij})$ and the observed surface
  brightness distributions $I_{\text{s}}(\bfx_{ij})$ for the classical
  and the smart detectors. The images have $101\times101$ pixels each
  and are based on the simulation with $N=10^6$ photon
  packages.}
\label{images.pdf}
\end{figure*}

\begin{figure*}
\centering
\includegraphics[width=0.245\textwidth]{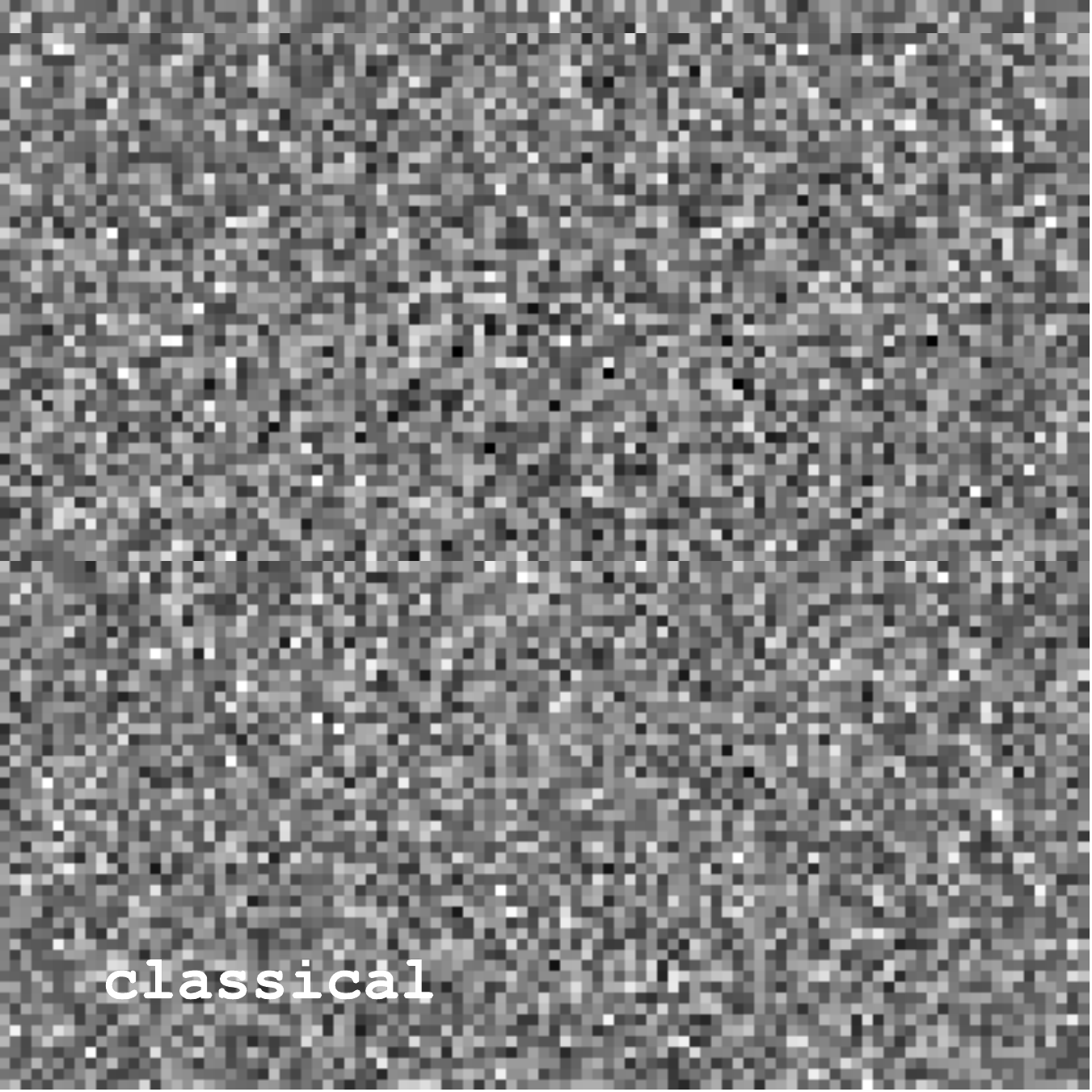}
\includegraphics[width=0.245\textwidth]{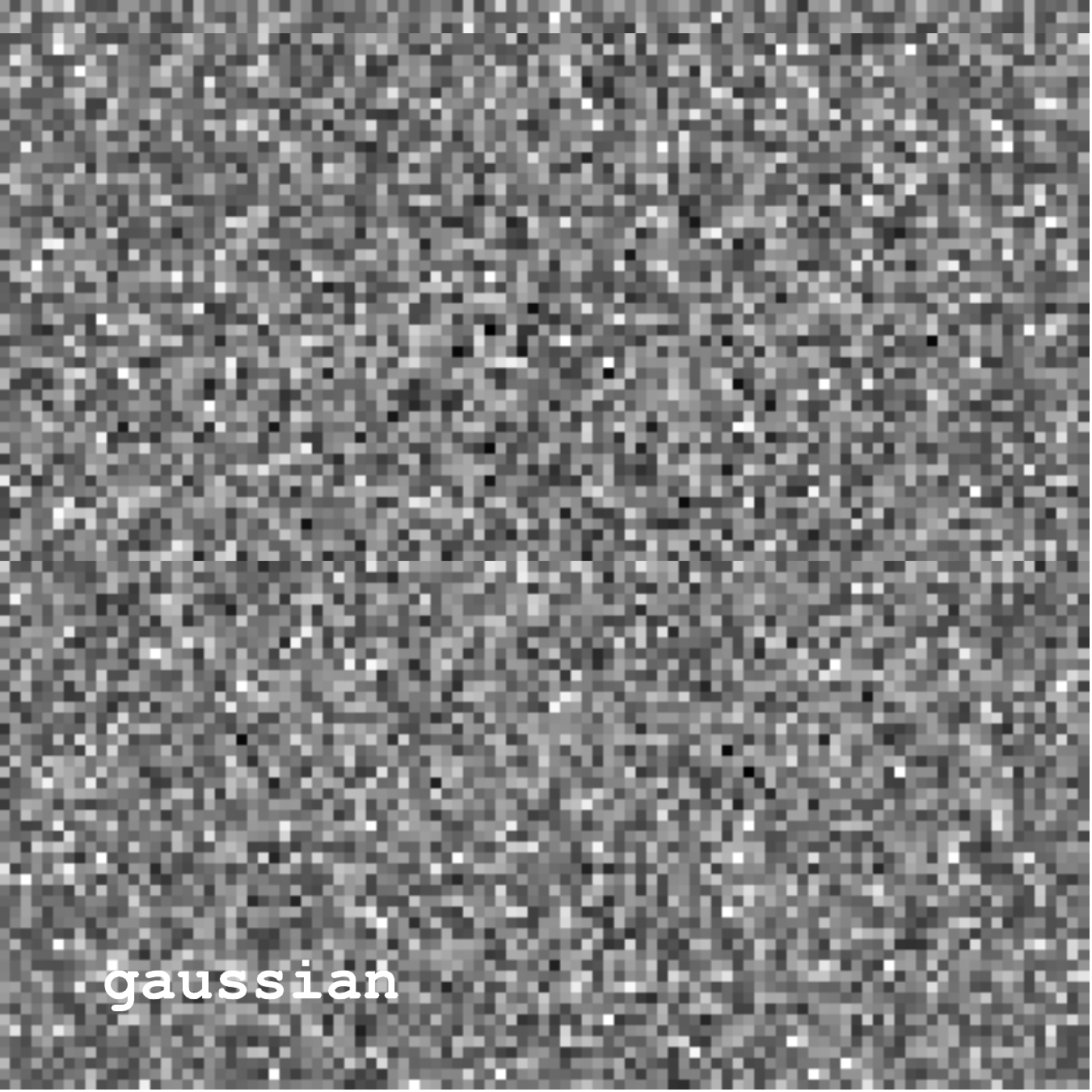}
\includegraphics[width=0.245\textwidth]{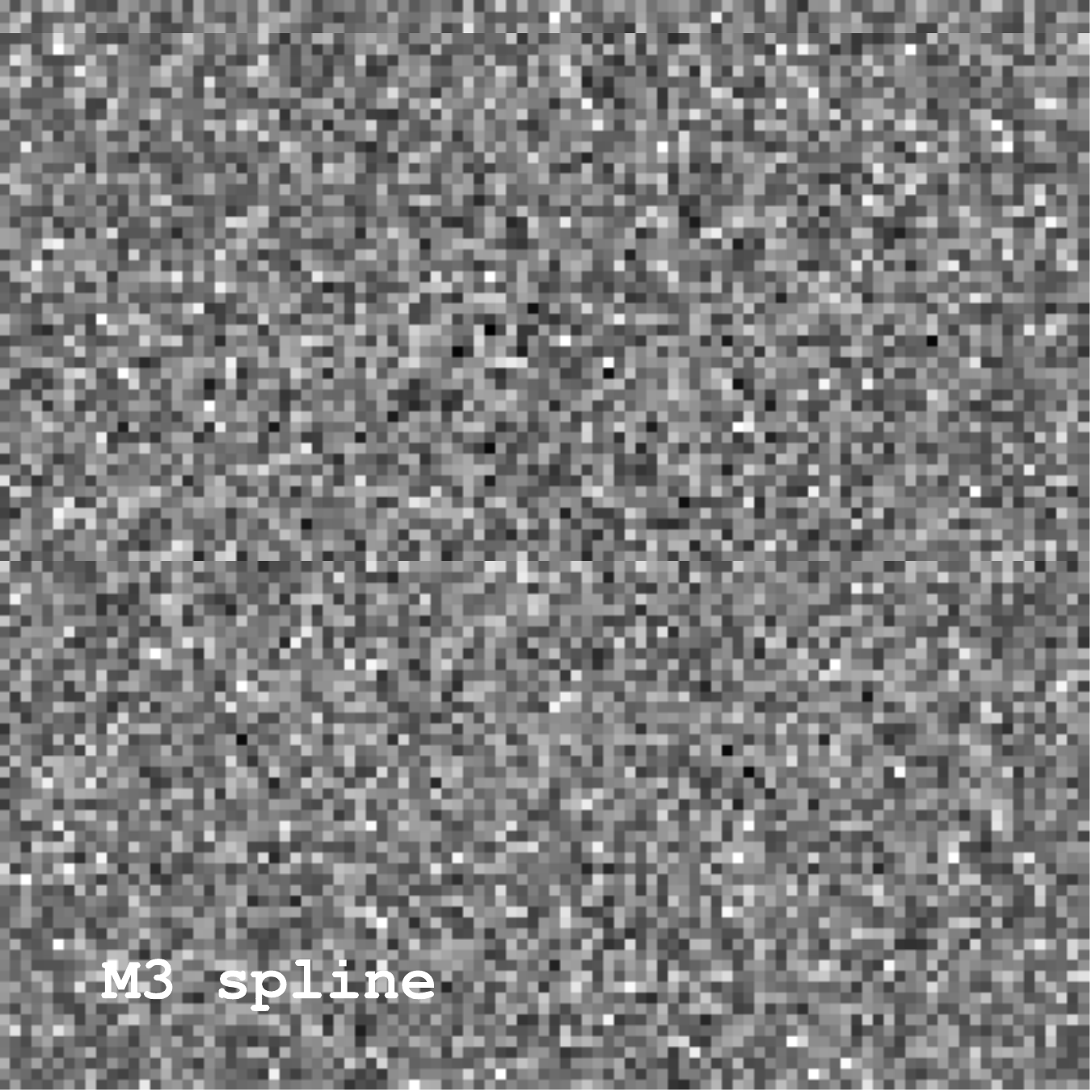}
\includegraphics[width=0.245\textwidth]{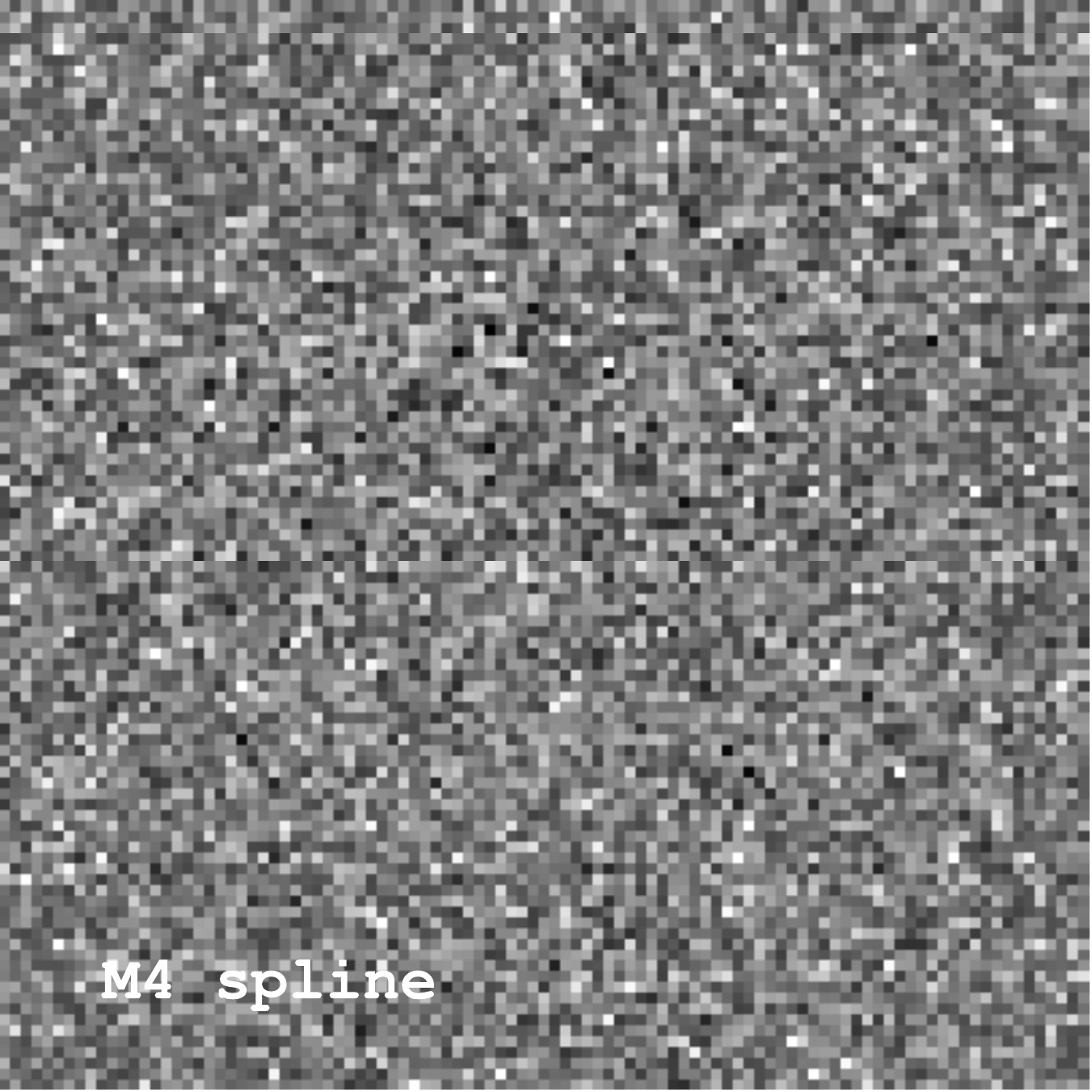} 
\caption{The noise distributions $\Delta I(\bfx_{ij})$ corresponding
  to the observed Plummer surface brightness distributions displayed
  in Figure~{\ref{images.pdf}}. }
\label{noise.pdf}
\end{figure*}

We tested the accuracy and the performance of our smart detectors
using two toy analytical surface brightness distributions on the plane
of the sky.  The first is a simple Plummer model, characterized by the
circularly symmetric surface brightness distribution
\begin{equation}
  I(\bfx)
  =
  \frac{L}{\pi b^2}
  \left(\frac{b^2}{b^2+R^2}\right)^{2},
\end{equation}
with $L$ the total luminosity and $b$ a scale parameter. We took the
value $b=10\Delta$, implying that the core of the model is well
resolved by the detector grid. The second model is an exponential
disc model, rotated over an angle of 20 degrees,
\begin{equation}
  I(\bfx)
  =
  \frac{L}{4\pi h_x h_y}\,
  \exp\left(-\frac{|x|}{h_x}-\frac{|y|}{h_y}\right),
\end{equation}
with $h_x$ and $h_y$ the scalelength and scaleheight
respectively. Choosing the disc parameters as $h_x=25\Delta$ and
$h_y=5\Delta$, we create a model with a relatively sharp edge and a
strong gradient in the surface brightness distribution.

We have used a classical detector and three smart detectors based on
the smoothing kernels~(\ref{gaussiankernel}), (\ref{M3splinekernel})
and~(\ref{M4splinekernel}), each of them with a total of
$101\times101$ grid points. For each model, we ran a set of Monte
Carlo simulations with the number of photon packages varying between $N=10^4$
and $N=10^8$, taking into account that each detector measures the same
set of photon packages (i.e.\ we use the same Monte Carlo realisation for
simulations with different detectors).

\begin{figure}
\centering
\includegraphics[width=0.45\textwidth]{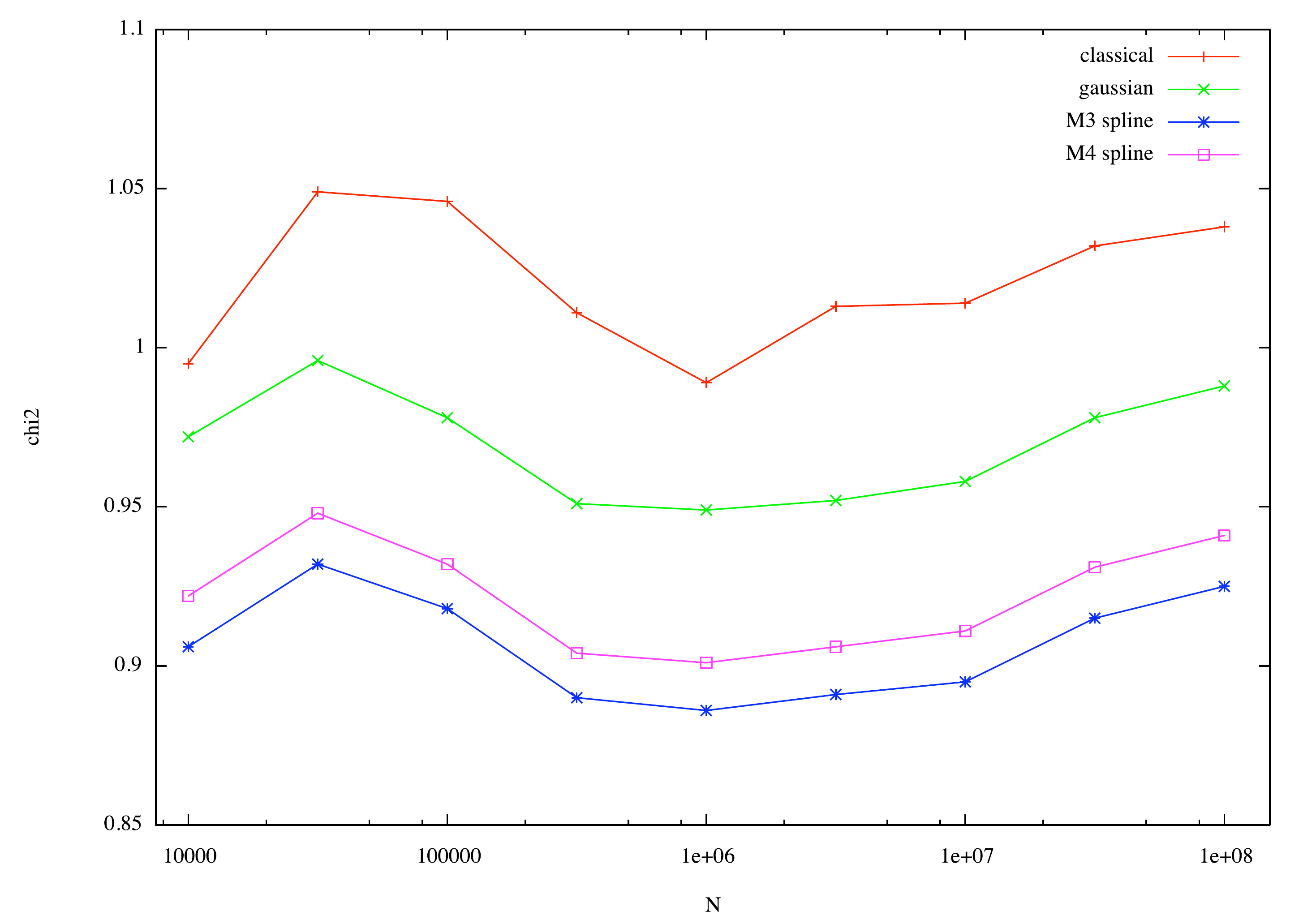}\hspace{2em}
\caption{The total noise parameter $\noise$, defined in
  equation~(\ref{chi2}), of the Plummer model for different detectors
  as a function of the total number of photon packages in the simulation. }
\label{plummerchi2.pdf}
\end{figure}

The different panels of figure~{\ref{images.pdf}} show the resulting
measured surface brightness distribution $I_{\text{s}}(\bfx)$ for the
various detectors for the simulations with $N=10^6$. Looking at this
set of images, it is immediately clear that the smart detectors manage
to qualitatively reproduce the surface brightness distribution
accurately. For a measure of the accuracy of the different detectors,
we consider the noise as the difference between the measured surface
brightness and the theoretical surface brightness in each pixel,
weighted by the expected Poisson noise $\sigma(x_{ij})$ in each pixel,
\begin{equation}
  \Delta I(\bfx_{ij})
  =
  \frac{I_{\text{s}}(\bfx_{ij})-I(\bfx_{ij})}{\sigma(\bfx_{ij})}.
\end{equation}
Figure~{\ref{noise.pdf}} show the noise images corresponding to the
Plummer model images from Figure~{\ref{images.pdf}}. It is clear that
these images are qualitatively very similar; thanks to the appropriate
choice of the smoothing lengths for the different kernels, there is no
correlation in the noise on a pixel-by-pixel scale. A quantitative
analysis of these noise images, however, shows that the level of the
noise in the smart detectors is suppressed. This is most easily
demonstrated using the total noise level, which we define as
\begin{equation}
  \noise
  =
  \frac{\sqrt{N}}{N_{\text{pix}}}
  \sum\left[  
    \frac{I_{\text{s}}(\bfx_{ij})-I(\bfx_{ij})}{\sigma(\bfx_{ij})}
  \right]^2,
\label{chi2}
\end{equation}
where the sum runs over all pixels. A factor $\sqrt{N}$ is included in
this formula to guarantee that the noise is asymptotically independent
of the total number $N$ of photon packages used in the
simulation. Figure~{\ref{plummerchi2.pdf}} shows the value of $\noise$
for the different detectors as a function of the total number of
photon packages in the simulation. This figure demonstrates that all smart
detectors reduce the noise compared to the classical detector. The
most efficient detector is the one based on the $M_3$ spline kernel;
for this detector the noise is reduced by about 10\%. For the $M_4$
spline kernel, the most popular finite support kernel in SPH
simulations, the noise reduction is slightly less efficient (about
8\%), whereas for the gaussian kernel the noise reduction is about
5\%.

\subsection{Origin of the noise reduction}

We have constructed our smart detectors based on two fundamental
changes applied to the classical detector smoothing
kernel~(\ref{Wheavi}), namely making it circularly symmetric and
choosing a smoothly decreasing function of radius. We can investigate
which of these two changes has the most important impact on the noise
reduction by constructing two new smoothing kernels in which only one
of these two changes is taken into account.

On the one hand, the circularly symmetric analogue of the classical
detector kernel~(\ref{Wheavi}) is readily found,
\begin{equation}
  W(R)
  =
  \frac{4}{\pi h^2}\,
  \left[
    H\left(R+\tfrac12h\right)-H\left(R-\tfrac12h\right)
  \right].
  \label{Cclassicalkernel}
\end{equation}
We find in the usual way $h_{\text{ref}}=2\Delta/\sqrt{3}$ as reference
smoothing length. On the other hand, the rectangular version of the
$M_3$ spline kernel~(\ref{M3splinekernel}) is
\begin{equation}
  W(\bfx)
  =
  \frac{1}{h^2}\,
  M_3\left(\frac{|x|}{h}\right)\,
  M_3\left(\frac{|y|}{h}\right),
  \label{RM3splinekernel}
\end{equation}
with
\begin{equation}
  M_3(u)
  =
  \begin{cases}
    \;\frac34-u^2
    &\quad
    \text{if $0\leq u\leq \tfrac12$},
    \\[0.5em]
    \;\tfrac12\,(\tfrac32-u)^2
    &\quad
    \text{if $\tfrac12\leq u\leq\tfrac32$},
    \\[0.5em]
    \;0
    &\quad
    \text{else}.
  \end{cases}
\end{equation}
For this kernel we obtain $h_{\text{ref}}=\Delta/\sqrt{3}$.

\begin{figure}
\centering
\includegraphics[width=0.45\textwidth]{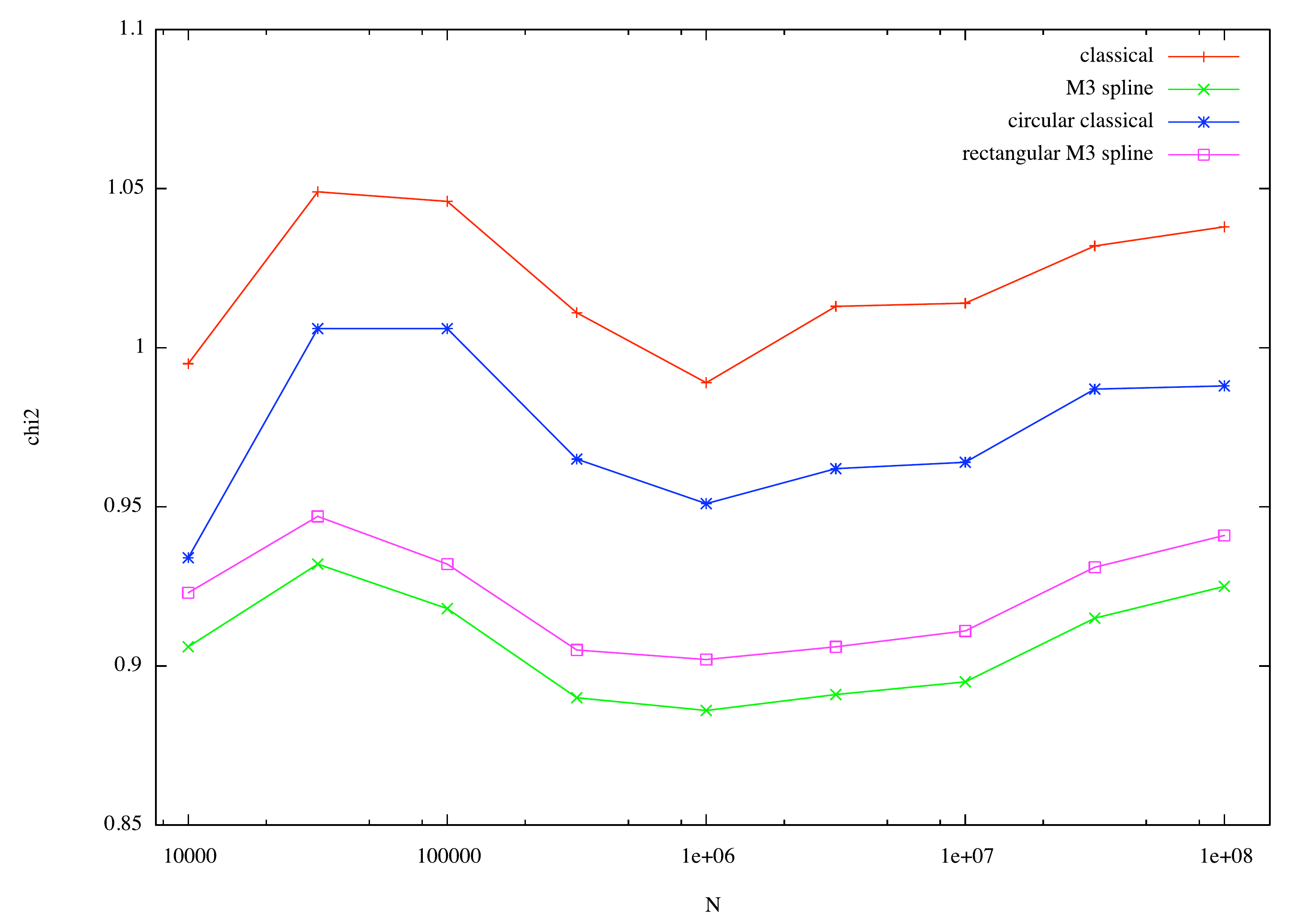}\hspace{2em}
\caption{The total noise parameter $\noise$ of the Plummer model for
  the rectangular and circular equivalents of the classical detector
  and the smart $M_3$ spline detector.}
\label{testchi2.pdf}
\end{figure}

In Figure~{\ref{testchi2.pdf}} we plot the total noise parameter
$\noise$ of these two new smart detectors in comparison with the
classical detector and the circularly symmetric $M_3$ spline
detector. Not surprisingly, we find that also these two new detectors
suppress the noise compared to the classical detector, but not as
efficiently as the $M_3$ spline detector. The noise reduction in a
detector based on the rectangular $M_3$ spline
kernel~(\ref{RM3splinekernel}) is more effective than the noise
reduction in the circular analogue~(\ref{Cclassicalkernel}) of the
classical detector. This means that applying a proper weight to the
photon packages according to their distance from the grid points is the most
important change if we want efficient noise reduction.

\section{Discussion and conclusions}

We have focused on an ill-studied area of Monte Carlo radiative
transfer simulations where a significant noise reduction can be
achieved, namely the detection of photon packages and the
corresponding construction of the observed surface brightness
distribution. The motivation of this work was the fact that the
classical detectors used in Monte Carlo simulations, while closely
mimicking a real CCD detector, do not use the full amount of
information that is available.

Based on the similarities between the construction of the surface
brightness distribution on a detector in Monte Carlo radiative
transfer simulations and the calculation of the density in SPH
hydrodynamical simulations, we have constructed a set of smart
detectors. These smart detectors improve on two aspects of the
classical detector: they assign the same weight to all photon packages
hitting the detector at the same distance from a grid point and they
give more weight to impacts close to a grid point than to impacts at
larger distances. We have tested different kinds of smart detectors,
based on different smoothing kernels frequently encountered in SPH
simulations, namely a gaussian kernel and two spline-based kernels
with finite support \citep{1981csup.book.....H,
  1985A&A...149..135M}. We have shown that these new detectors, while
preserving the same effective resolution, reduce the noise in the
surface brightness distributions compared to the classical
detectors. It is demonstrated that the lion's share of this noise
reduction is due to a proper weighing of the photon packages with the
distance between the impact location and the grid point.

The most efficient smart detector is found to be a detector based on
the $M_3$ spline kernel, for which the noise reduction amounts to some
10\%. While this might seem a modest improvement, one should take into
account that the noise reduction in Monte Carlo simulation goes as
$1/\sqrt{N}$. A reduction of the noise with 10\% is hence equivalent
to a reduction of the number of photon packages with 20\%, which is a
significant improvement. Moreover, it should be stressed that this
noise reduction basically comes for free, since the practical
implementation of the smart detectors is straightforward and the
additional computational cost is completely negligible. We hence
strongly recommend the use of smart detectors in Monte Carlo radiative
transfer simulations.

The link with SPH simulations might stimulate to look for further ways
to optimize the estimates of the surface brightness distribution. One
major difference between our current problem and the interpolation in
SPH simulations is that we use a fixed smoothing length, whereas SPH
simulations typically use a spatially (and temporally) varying
smoothing length. This way it is possible to take full advantage of
the particle distribution to resolve local density structures. Each
particle in an SPH simulation typically has an individual smoothing
length which is fine-tuned such that each particle interacts with a
similar number of neighbors. Unfortunately, it seems hard to introduce
variable smoothing lengths here in our current Monte Carlo radiative
transfer case. The main difference is that the smoothing procedure in
SPH is executed when all particle positions are known, whereas in
Monte Carlo radiative transfer the photon packages gradually hit the
detector and they must be smoothed out onto the detector before it is
known where the other photon packages will arrive. It is impossible to
know {\em{a priori}} how many neighbors a photon package will
ultimately have and hence to adapt the smoothing length on an
individual basis. One could potentially do a test run with a limited
number of photon packages to obtain a first crude estimate of the
expected surface brightness distribution and adapt the smoothing
lengths accordingly. However, a more elegant option seems to apply an
adaptive filtering to the simulated images {\em{after}} the
simulation. Several approaches have been developed for this goal such
as wavelet-based algorithms \citep{1998A&AS..128..397S}, adaptive
binning techniques \citep{1998A&AS..128..397S, 2003MNRAS.342..345C} or
adaptive kernel smoothing techniques \citep{1991AN....312..345R,
  1996ApJ...461..622H, 2006MNRAS.368...65E}.

\end{document}